\documentclass[%
 reprint,
 superscriptaddress,
 amsmath,amssymb,
 aps,
 pra,
floatfix,
]{revtex4-2}
\usepackage{graphicx}
\usepackage{dcolumn}
\usepackage{bm}
\usepackage{physics}
\usepackage{color}
\usepackage{xcolor}
\usepackage[]{comment}

\begin{document}

\preprint{APS/123-QED}

\title{Rate-fidelity trade-off in cavity-based remote entanglement generation}

\author{Kazufumi Tanji}
 \email{kizehemu@keio.jp}

\affiliation{%
 Department of Electronics and Electrical Engineering, Faculty of Science and Engineering, Keio University, 3-14-1 Hiyoshi, Kohoku-ku, Yokohama-shi, Kanagawa, 223-8522, Japan
}%

\author{Hiroki Takahashi}
\affiliation{
 Experimental Quantum Information Physics Unit, Okinawa Institute of Science and Technology Graduate University, 1919-1 Tancha, Onna, Kunigami, Okinawa 904-0495, Japan
}%

\author{Wojciech Roga}

\author{Masahiro Takeoka}
\affiliation{%
 Department of Electronics and Electrical Engineering, Faculty of Science and Engineering, Keio University, 3-14-1 Hiyoshi, Kohoku-ku, Yokohama-shi, Kanagawa, 223-8522, Japan
}%

\date{\today}

\begin{abstract}
The qubit scalability imposes a paramount challenge in the field of quantum computing.
Photonic interconnects between distinct quantum computing modules provide a solution to deal with this issue.
The fundamental part of this approach is entanglement distribution via travelling photons emitted by matter qubits. However, randomness of the spontaneous emission in the matter qubits limits both the entanglement fidelity and the generation rate. 
In this paper, by numerical and analytical methods, we investigate the relationship between the entanglement affected by the spontaneous emission and the waveform of the pump pulse used in the photon generation. 
We confirm and analyze a rate-fidelity trade-off in the entanglement swapping with Gaussian pump pulses and show that a simple extension to non-Gaussian pump pulses improves the trade-off in a certain parameter region.
Furthermore we extend our analysis to entanglement distribution in the general multipartite setting and show that the analysis of the bipartite entanglement can be straightforwardly applied in this case as well. 
\end{abstract}

\maketitle


\section{Introduction}\label{section1}

Quantum computers have the potential to solve classically intractable problems~\cite{Shor1994,Grover1996}. However, in comparison to the requirement of such applications, the current number of qubits in any physical system~\cite{Kim2023, Moses2023, Schymik2022} is limited by various physical constraints.
A promising solution to this scalability problem is distributed quantum computing~\cite{Kimble2008, Wehner2018,Monroe2014,Henriet2020}, in which small quantum computers connected via a photonic network compose a large-scale quantum computer.

\begin{figure}
    \includegraphics[width=8.6truecm]{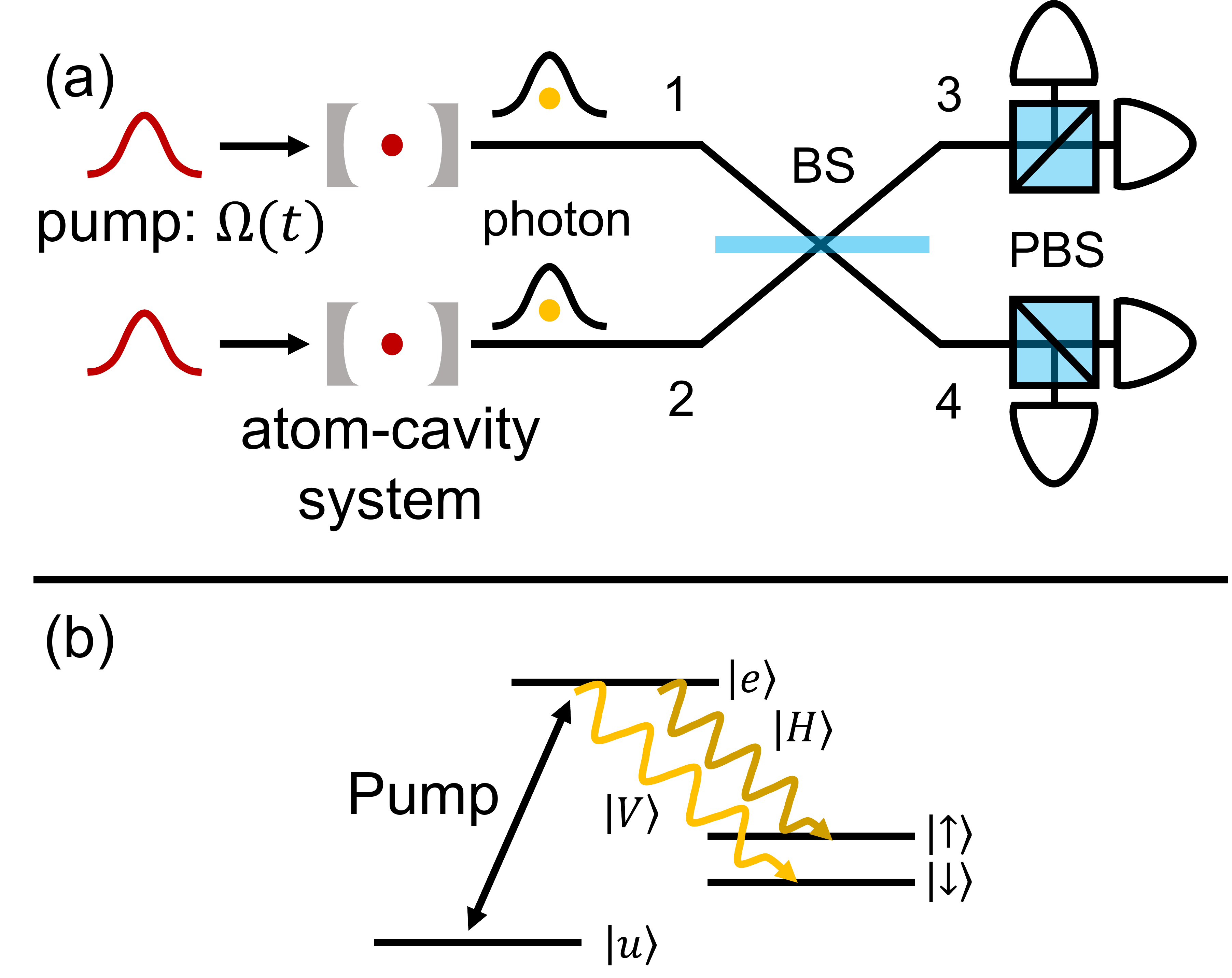}
    \caption{(a) The model of two-photon entanglement swapping~\cite{Duan2003}. The internal states of atoms and polarization of a photon are entangled. The measurement unit is composed of beam splitter (BS), polarization beam splitters (PBS), and single photon detectors. (b) The atomic energy structure. The polarization of photon correlates the spin state of the ions.}
    \label{fig:fig1}
\end{figure}

Fig.~\ref{fig:fig1}(a) shows an elementary structure of the photonic interconnects, where two atom-cavity systems emit photons entangled with the internal degrees of freedom of the atoms, photons interfere at the beam splitter, and finally are detected, realizing the two-photon entanglement swapping~\cite{Duan2003}. 
Note that in this paper the``atom" refers to a physical system that hosts a stationary qubit including not only natural atoms such as neutral atoms and atomic ions but also artificial atoms. 
The atom-cavity system is widely used in superconducting circuits~\cite{Blais2021}, neutral atoms~\cite{Reiserer2015}, trapped ions~\cite{Reiserer2015}, quantum dots~\cite{Najer2019}, and diamond color centers~\cite{Janitz2020} due to the enhancement of  the photon emission rate by the cavity~\cite{Purcell1946}. 
Note that two-photon entanglement swapping is experimentally more advantageous than the one-photon entanglement swapping~\cite{Cabrillo1999} since the two-photon method relaxes the condition of phase stabilization in the photonic links~\cite{Monroe2014}. 
Such entanglement swapping has been experimentally demonstrated in superconducting circuits~\cite{Narla2016}, neutral atoms~\cite{Hofmann2012, Rosenfeld2017, VanLeent2022}, trapped ions~\cite{Hucul2015, Stephenson2020, Krutyanskiy2023}, quantum dots~\cite{Zopf2019, BassoBasset2019}, and diamond color centers~\cite{Bernien2013}. 
In those experiments, the performance of the entanglement swapping was limited by not only technical issues but also an inherent problem of the spontaneous emission.

Fig.~\ref{fig:fig1}(b) illustrates a typical model of the atomic energy structure. The atom in the initial state $|u\rangle$ is excited by an external pump field resonant on the $\ket{u}\leftrightarrow\ket{e}$ transition, which subsequently induces an emission of polarized photons $\ket{H}$ and $\ket{V}$ into the cavity while accompanying the atomic state transitions to $\ket{\downarrow}$ and $\ket{\uparrow}$ respectively (More details will be given in the following sections). 
During this process, two kinds of spontaneous emission may occur: one is a decay to the initial state ($\gamma_u$); the other, a decay to other states ($\gamma_g$). In the former case the spontaneous emission can be followed by re-excitement of the atom but in the latter the atom is no longer coupled to the pump field and the re-excitement does not occur. 
Note that if the atom-cavity coupling is sufficiently large, vacuum-stimulated Raman adiabatic passage (vSTIRAP) can adiabatically eliminate the effects of these spontaneous emissions from the excited level~\cite{Kuhn1999,AxelKuhn2002}.
In some cases, however, it is not easy to experimentally achieve sufficiently strong coupling~\cite{Takahashi2020, Janitz2020}, and entanglement swapping has not been thoroughly investigated in such an intermediate coupling regime.

The atomic decay characterized by $\gamma_u$ may lead to new emissions of photons through the re-excitation of the atom which results in the output photonic state described by a statistical mixture.
This degrades the two-photon interference visibility and thus the fidelity of the entangled atoms after the swapping. Such an effect has been confirmed both in theory~\cite{Gao2020} and experiments~\cite{Walker2020,Meraner2020}. 
Regarding the decay characterized by $\gamma_g$, once it happens, the atom can no longer be re-excited and thus this decay contributes only to the decrease of the photon emission probability~\cite{Vasilev2010,Goto2019,Utsugi2022}. 

There are two approaches to circumvent the adversarial effects caused by the spontaneous emission associated with $\gamma_u$. 
One is choosing energy levels such that $\gamma_u$ is small. This allows for obtaining high visibility photons~\cite{Walker2020}, while it has a drawback that it requires a long time for the ground state preparation with optical pumping since a small $\gamma_u$ implies a weak dipole moment of the corresponding transition.
In consequence, the repetition rate for the photon generation would be limited. 
Another approach is to optimize the pump pulse waveform. There have been theoretical studies focused on optimization of the pump pulse shape to maximize the photon generation rate~\cite{Vasilev2010,Gao2020,Utsugi2022}. Among them Refs.~\cite{Vasilev2010,Utsugi2022} considered the situation with negligible $\gamma_u$, while in~\cite{Gao2020} the authors suggested that, for finite $\gamma_u$, the optimal pump pulse should deviate from a typical Gaussian waveform. 
However, comprehensive analysis of the role of the pump pulse shape in remote entanglement generation that takes into account both generation rate and fidelity is still missing.

In this paper, we numerically and analytically study the entanglement generation between two distant atom-cavity systems through photonic interconnects affected by spontaneous emissions of the atoms followed by re-excitation. 
First, we consider Gaussian-shaped pump pulses with various pulse areas and pulse widths, and numerically show quantitative trade-off between the entanglement generation rate and the fidelity. 
The pump pulse shape determines if the excitation is adiabatic or not. 
For various parameter regions, we clarify which shape allows for the saturation of the trade-off with both adiabatic and non-adiabatic excitation.
In addition, we explore the possibility to beat the trade-off by employing a non-Gaussian waveform for the pump pulse. Our numerical study strongly suggests that it is possible to beat the Gaussian pulse trade-off with asymmetric Gaussian pulses even though its benefit is not substantial, a few percent increase in the fidelity. 
We also derive an analytical expression for the fidelity in the limit of high entanglement generation rate. 
Combining it with numerical analysis, we give an approximated but simple expression that is solely a function of the cooperativity of the atom-cavity system. 
Finally, we consider the extension of our scenario such that more than two distant atoms are connected by photonic interconnects to share multi-partite entanglement. We show that state fidelities in any multi-partite entanglement generation in this scenario can be derived from only two-photon correlation functions. 
This result simplifies the future analyses on quantum networks consisting of photonically interconnected atoms. 

The paper is organized as follows. 
Section II describes the model of the system and necessary theoretical tools. Numerical and analytical results are shown in Sec.~III. We also discuss the extension of our model to the multi-partite entanglement generation among more than two distant atoms in Sec.~IV. Section IV concludes the paper.

\section{Model and Methods}
\label{section2}
In this section, we describe our model and theory. In Sec.~\ref{subsection2.1}, we derive the relationship between the atom-atom bipartite entanglement fidelity and the waveform of the photons in our model. 
The atom-cavity system is described in Sec.~\ref{subsection2.3} and corresponding quantum master equations and the quantum regression theorem (QRT) are derived in Sec.~\ref{subsection2.4}. 

\subsection{Entanglement swapping between two nodes and the waveforms of the photons}\label{subsection2.1}

Fig.~\ref{fig:fig1}(a) illustrates the photonic interconnect of atom-cavity systems via entanglement swapping. 
Each atom-cavity system emits a polarized photon that is entangled with the respective atom. 
The state of the system before the beam splitter is given by 
\begin{eqnarray}
    \label{eq:eq1}
        \ket{\Psi}&=&\ket{\psi_1}\ket{\psi_2} ,\nonumber\\
        \ket{\psi_i}&=&\frac{1}{\sqrt{2}}\left(a_{p,i}^{H\dagger}\ket{\uparrow}_i\ket{0}_{p,i}-a_{p,i}^{V\dagger}\ket{\downarrow}_i\ket{0}_{p,i}\right).
\end{eqnarray}
where $\ket{\psi_i} (i=1,2)$ is the wavefunction at the spatial mode $i$ and $\ket{\uparrow}_i$ and $\ket{\downarrow}_i$ denote the atomic internal states. 
The creation operator of the propagating photons $a_{p,i}^{\mu\dagger} (\mu=H,V)$ characterizes each spatial and polarization modes, where $H$ and $V$ represent the horizontal and vertical polarizations, respectively.
To take into account the time distribution of the photons, we represent the creation operator as 
\begin{equation}
    \label{eq:eq2}
    a_{p,i}^{\mu\dagger}\ket{0}\rightarrow\int_{0}^{\infty}dt \phi_i^{\mu *}(t)a_{p,i}^{\mu\dagger}(t)\ket{0},
\end{equation}
where $\phi(t)$ is the waveform of the photon, which satisfies the normalization condition, $\int_0^\infty \abs{\phi(t)}^2=1$. 

Next, the photons interfere with each other on the beam splitter, pass polarization beam splitters and are detected at photon detectors. The system of beam splitters and detectors realizes the Bell measurement. 
Suppose photons are detected at time $t$ and $t'$ in mode 3 (Fig.~\ref{fig:fig1}) with horizontal and vertical polarizations, respectively. Then the remaining state of atoms is given by  
\begin{equation}
    \label{eq:eq3}
    \ket{\Psi_{\rm ion}(t,t')}=\frac{1}{\sqrt{\mathcal{N}}}\bra{0}a_{p,3}^H(t)a_{p,3}^V(t')\ket{\Psi_{\rm out}}, 
\end{equation}
where $\mathcal{N}$ is a normalization factor, and $\ket{\Psi_{\rm out}}$ is the state after the beamsplitter (see Eq.~(\ref{eq:eq.A.3}) in Appendix A). 
In the realistic situation, however, it is difficult to resolve the detection times. Then, the final atomic state is given in a density matrix form where all trajectories, i.e. the states with different $t$ and $t'$, are integrated. 
As a consequence, the fidelity between $\ket{\Psi_{\rm Bell}^+}=\frac{1}{\sqrt{2}}(\ket{\uparrow\downarrow}+\ket{\downarrow\uparrow})$ and the generated atom-atom entanglement after a successful Bell measurement is described as 
\begin{equation}
    \label{eq:eq4}
        F=\frac{1}{2}(1+\Re{J}), 
\end{equation}
where 
\begin{equation}
\label{eq:eq5}
        J=\int dt \phi_1^{H*}(t)\phi_2^{H}(t)\int dt' \phi_1^{V}(t')\phi_2^{V*}(t').
\end{equation}
Thus, the fidelity is proportional to the real part of the correlation function $J$ of the photon waveforms. 
See Appendix~\ref{AppendixA} for the details of the derivation.
As mentioned in the previous section, since the spontaneous decay may cause re-excitation of the atom, a statistical mixture of different photon emission waveforms should be taken into account.
Therefore, we need to consider an average of these waveforms in $J$: 
\begin{equation}
    \label{eq:eq6}
    \langle J\rangle=\left\langle\int dt \phi_1^{H*}(t)\phi_2^{H}(t)\int dt' \phi_1^{V}(t')\phi_2^{V*}(t')\right\rangle,
\end{equation}
where $\langle \, \rangle$ stands for the statistical average.
The waveform of each photon and the effect of re-excitation depends on the dynamics in the atom-cavity system and is discussed in Sec.~\ref{subsection2.3} and~\ref{subsection2.4}

\subsection{Model for the atom-cavity system}\label{subsection2.3}
\begin{figure}[htbp]
    \centering
    \includegraphics[width=8.6truecm]{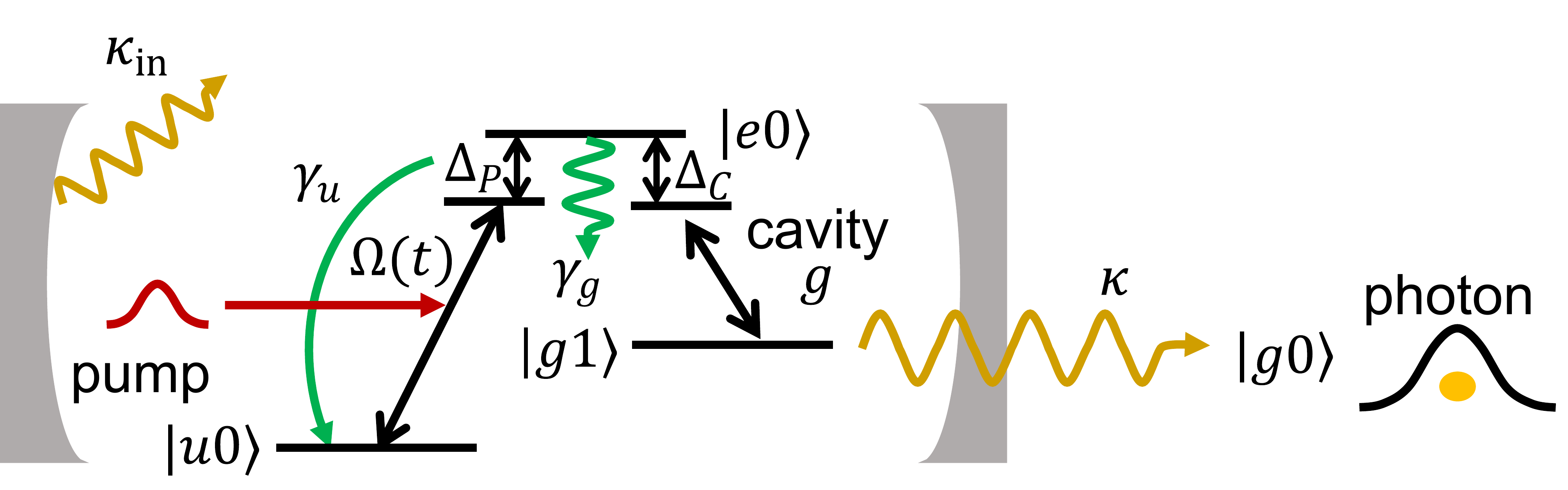}
    \caption{The model for the atom-cavity system. 
    $\Omega(t)$: the Rabi frequency of the driving pulse, $\Delta_{u},\Delta_{p}$: detunings, $g$: the atom-cavity coupling, $\kappa$: the cavity decay rate, $\gamma_u$: the re-excitable decay, and $\gamma_g$: the non-re-excitable decay.}
    \label{fig:Fig3}
\end{figure}
In this subsection, we describe the atom-cavity system and the derivation of $\langle J \rangle$ based on it. 
The atom-cavity model considered here is shown in Fig.~\ref{fig:Fig3}.
The ket vector represents the atomic and cavity photon states, for example, $|u0\rangle=|u\rangle_{\rm atom} |0\rangle_{\rm photon}$ means that the atom is in the state $|u\rangle$ and there is no photon in the cavity.  
The system consists of a three level atom ($|u\rangle$, $|e\rangle$, and $|g\rangle$) and the transition between $|e\rangle$ and $|g\rangle$ is coupled to the cavity mode with the coupling $g$.
$\gamma_u$ is the spontaneous decay rate from $|e\rangle$ to $|u\rangle$ and the total decay rate to the other states is denoted as $\gamma_g$. 
The photons of the cavity are emitted to the outside of the cavity with a decay rate $\kappa$. 
The system is initialized in $|u0\rangle$ and then coupled to $|e0\rangle$ by the pump pulse with a Rabi frequency of $\Omega(t)$. $\Delta_p$ and $\Delta_c$ are the detunings between the atomic transition frequency and the pump pulse frequency and the cavity resonance, respectively, as shown in Fig.~\ref{fig:Fig3}. 

For simplicity and to save computational time, in our numerical analysis we assume
\begin{equation}
    \label{eq:eq13}
        \gamma_g=\Delta_p=\Delta_c=0. 
\end{equation}
Note that if $\gamma_g$ is finite, it simply reduces the probability of photon emission~\cite{Goto2019}.  
Also, finite detunings contribute to suppress the spontaneous emission in principle, while it cannot fully remove the effect of the spontaneous emission as pointed out in~\cite{Walker2020}. Therefore, our assumption does not change the analysis qualitatively.

Moreover, Fig.~\ref{fig:Fig3} is a simplified model in which we assume that the ground state Zeeman splitting is small, i.e. $|\uparrow\rangle$ and $|\downarrow\rangle$ are degenerate in $|g\rangle$. 
Thus, we assume that the waveform of the propagating photons are the same for both polarizations, i.e.  $\phi_i^H(t)=\phi_i^V(t)$.
Then we rewrite Eq.~(\ref{eq:eq6}) as follows,
\begin{equation}
    \label{eq:eq14}
    \langle J \rangle = \iint dtdt' \langle \phi_1^*(t)\phi_1(t') \rangle\langle \phi_2(t)\phi_2^*(t') \rangle,
\end{equation}
where we used the linearity of the average and independence between mode 1 and mode 2, and omitted the superscripts for polarization. The average $\langle \phi_i^*(t)\phi_i(t') \rangle \quad (i=1,2)$ is equivalent to the first-order coherence of the field. Then, the correlation function is represented by the first-order coherence as
\begin{equation}
    \label{eq:eq15}
    \langle J \rangle = \iint dtdt' \langle a_{p,1}^\dagger(t)a_{p,1}(t') \rangle\langle a_{p,2}(t)a_{p,2}^\dagger(t') \rangle,
\end{equation}
where $a_{p,i}^\dagger$ and $a_{p,i}$ ($i=1,2$) are creation and annihilation operators for the propagating photons in the $i$-th mode, respectively, and the brackets for the creation and annihilation operators in the right-hand side stands for the expectation value in the standard quantum mechanics notation.
The propagating photons and the cavity photons are connected through the input-output relation,
\begin{equation}
    \label{eq:eq16}
    -i\sqrt{2\kappa}a_c(t)=a_p(t),
\end{equation}
where $a_c(t)$ is the annihilation operator for the cavity field. 
Note that $\phi_i(t)$ is not normalized in general since the photon is not always emitted. 
On the other hand, in Sec.~IIA and B, $\phi_i(t)$ is always normalized since it represents the waveform of the conditional state when the Bell measurement signals a success.  
Therefore, we always use this normalization when we apply the results from this subsection to those in Sec.~IIA and B. 

Finally, the correlation function is given by 
\begin{equation}
    \label{eq:eq17}
    \langle J \rangle = \frac{4\kappa^2\iint dtdt' \langle a_{\mathrm{c},1}^\dagger(t)a_{\mathrm{c},1}(t') \rangle\langle a_{\mathrm{c},2}(t)a_{\mathrm{c},2}^\dagger(t') \rangle}{P_{\rm ex,1}P_{\rm ex,2}},
\end{equation}
where $P_{\rm ex,1}$ and $P_{\rm ex,2}$ are the probabilities of emitting a photon in mode 1 and 2, respectively. 
Similar expressions were obtained in~\cite{Gao2020,Fischer2016}.
The probability $P_{\rm ex,i}$ can be calculated from the Lindblad quantum Master equation and the two-point correlation function $a_{c,i}^\dagger(t)a_{c,i}(t')$ by using the quantum regression theorem, see Sec.~\ref{subsection2.4}. 
If the photons from the two atom-cavity systems have exactly identical properties, Eq.~(\ref{eq:eq17}) is simplified to 
\begin{equation}
    \label{eq:eq18}
    \langle J \rangle = \frac{4\kappa^2\iint dtdt' \abs{\langle a_{\mathrm{c}}^\dagger(t)a_{\mathrm{c}}(t') \rangle}^2}{P_{ex}^2}.
\end{equation}

\subsection{Master equation and the quantum regression theorem}\label{subsection2.4}
The quantities from the above model are calculated by using the master equation and quantum regression theorem (QRT). 
Under the rotating-wave approximation, the Lindblad master equation of our model is given by 
\begin{eqnarray}
        \label{eq:eq19}
    \frac{d}{dt}\rho(t)&=&-i\left(H_\mathrm{eff}(t)\rho(t)-\rho(t)H_\mathrm{eff}^\dagger(t)\right)\nonumber\\
    &&+2\gamma_u\ketbra{u}{e}\rho(t)\ketbra{e}{u}\nonumber\\
    &&+2\kappa a_c\rho(t)a_c^{\dagger}, 
\end{eqnarray}
where
\begin{eqnarray}
    \label{eq:eq20}
    H_\mathrm{eff}(t)&=&-i\kappa a_c^\dagger a_c-i\gamma_u\ketbra{e}{e}\nonumber\\
    &&+\left[\Omega(t)\ketbra{u}{e}+ga_c\ketbra{e}{g}+h.c.\right],
\end{eqnarray}
and $\gamma_u$ is the spontaneous decay rate from $\ket{e}$ to $\ket{u}$, $\kappa$ the cavity decay rate, $\Omega(t)$ the Rabi frequency of the pump, $g$ the coupling between the atom and the cavity, as shown in Fig.~\ref{fig:Fig3}, and $h.c.$ denotes Hermitian conjugation.
The photon emission probability $P_{\rm ex}$ reads 
\begin{equation}
    \label{eq:eq21}
        P_{\rm ex}=2\kappa\int dt\langle a_c^{\dagger}(t)a_c(t)\rangle
        =2\kappa\int dt \bra{g1}\rho(t)\ket{g1},
\end{equation}
which is obtained by solving the master equation. 

To compute the two-point correlation function we use the QRT~\cite{CrispinGardiner2004},
\begin{eqnarray}
    \label{eq:eq22}
        \langle a_c^\dagger(t)a_c(t')\rangle&=&\Tr[\Lambda(t,t')a_c],\\
    \label{eq:eq23}
        \frac{d}{dt'}\Lambda(t,t')&=&-i\left(H_\mathrm{eff}(t')\Lambda(t,t')-\Lambda(t,t')H_\mathrm{eff}^\dagger(t')\right)\nonumber\\
    &&+2\gamma_u\ketbra{u}{e}\Lambda(t,t')\ketbra{e}{u}\nonumber\\
    &&+2\kappa a_c\Lambda(t,t')a_c^{\dagger},\\
    \label{eq:eq24}
    \Lambda(t,t)&=&\rho(t)a_c^\dagger.
\end{eqnarray}
Note that although $t'\geq t$ in the above formalism, 
one can find $\langle a_c^\dagger(t)a_c(t')\rangle$ for arbitrary $t$ and $t'$
due to the symmetry of the two-point correlation function: 
\begin{equation}
    \label{eq:eq25}
    \langle a_c^\dagger(t)a_c(t')\rangle^*=\langle a_c^\dagger(t')a_c(t)\rangle,
\end{equation}
which is useful to compute $J$, for example, in Eq.~(\ref{eq:eq18}). 

When the system initially populates $\ketbra{u0}{u0}$, the density matrix at an arbitrary time is reduced to a $4\times4$ matrix spanned by $\{\ket{u0}, \ket{g1}, \ket{e0}, \ket{g0} \}$. 
Then, the matrix form of the Hamiltonian and the density matrix are given by
\begin{align}
    H_\mathrm{eff}(t)&=\begin{bmatrix}
    0 & 0 & \Omega(t) & 0\\
    0 & -i\kappa & g & 0\\
    \Omega^*(t) & g & -i\gamma_u & 0\\
    0 & 0 & 0 & 0\\
\end{bmatrix},\label{eq:eq26}\\
    \rho(t)&=
    \begin{bmatrix}
        \rho_{11}&\rho_{12}&\rho_{13}&0\\
        \rho_{12}^*&\rho_{22}&\rho_{23}&0\\
        \rho_{13}^*&\rho_{23}^*&\rho_{33}&0\\
        0&0&0&\rho_{44}\\
    \end{bmatrix}\label{eq:eq27},
\end{align}
respectively.

With the same initial condition, $\Lambda(t,t)$ (Eq.~(\ref{eq:eq24})) is expressed as
\begin{equation}
\label{eq:eq28}
    \Lambda(t,t)=\rho(t)a_c^\dagger=\begin{bmatrix}
            0&0&0&\rho_{12}(t)\\
            0&0&0&\rho_{22}(t)\\
            0&0&0&\rho_{23}^*(t)\\
            0&0&0&0\\
        \end{bmatrix}.
\end{equation}
Using Eq.~(\ref{eq:eq28}) as the initial state of Eq.~(\ref{eq:eq23}), we find that the non-zero elements of $\Lambda(t,t')$ are in the same matrix entries as the non zero entries of $\Lambda(t,t)$, i.e.  
\begin{equation}
\label{eq:eq29}
    \Lambda(t,t')=\begin{bmatrix}
        0&0&0&\lambda_1(t,t')\\
        0&0&0&\lambda_2(t,t')\\
        0&0&0&\lambda_3(t,t')\\
        0&0&0&0\\
    \end{bmatrix}.
\end{equation}
This fact allows us to simplify Eq.~(\ref{eq:eq23}) to the following form: 
\begin{equation}
        \frac{d}{dt'}\ket{\lambda(t,t')}=-iH_{3\times3}(t)
    \ket{\lambda(t,t')}\label{eq:eq30},
\end{equation}
where $H_{3\times3}$ is the upper-left $3\times3$ matrix of $H_\mathrm{eff}$, and
\begin{equation}
    \label{eq:eq31}
    \ket{\lambda(t,t')}=\begin{bmatrix}
        \lambda_1(t,t')\\
        \lambda_2(t,t')\\
        \lambda_3(t,t')
    \end{bmatrix}.
\end{equation}
Eq.~(\ref{eq:eq30}) is in fact a three-dimensional Schr\"odinger equation.
Finally, the two-point correlation function is
\begin{equation}
    \label{eq:eq32}
    \langle a_c^\dagger(t)a_c(t')\rangle=\Tr[\Lambda(t,t')a_c]=\lambda_2(t,t').
\end{equation}
Therefore, the fidelity is obtained by solving the master equation just once and three-dimensional Schr\"odinger equation sufficient number of times for the integration in Eq.~(\ref{eq:eq18}).

\section{Results}\label{section3}
\subsection{Numerical results}\label{subsection3.1}

\begin{table}
    \caption{Parameters for the numerical simulation normalized by decay rate $\gamma_u$. $C$ is the cooperativity, $C=\frac{g^2}{\kappa\gamma_u}$.}
    \begin{ruledtabular}
    \label{table:Tab1}
    \begin{tabular}{ccccc}
    Regime &$g/\gamma_u$ & $\kappa/\gamma_u$ & $C$\\ \hline
    Intermediate (a)&1 & 1 & 1\\
    Strong (b)&$\sqrt{10}$ & 1 & 10\\
    Weak (c) & $1/\sqrt{10}$ & 1 & 0.1\\
    Purcell (d)&5 & 25 & 1\\
    Lossy atom (e)&0.2 & 0.04 & 1\\
\end{tabular}
\end{ruledtabular}
\end{table}

\begin{figure*}[htbp]
    \includegraphics[width=13.5truecm]{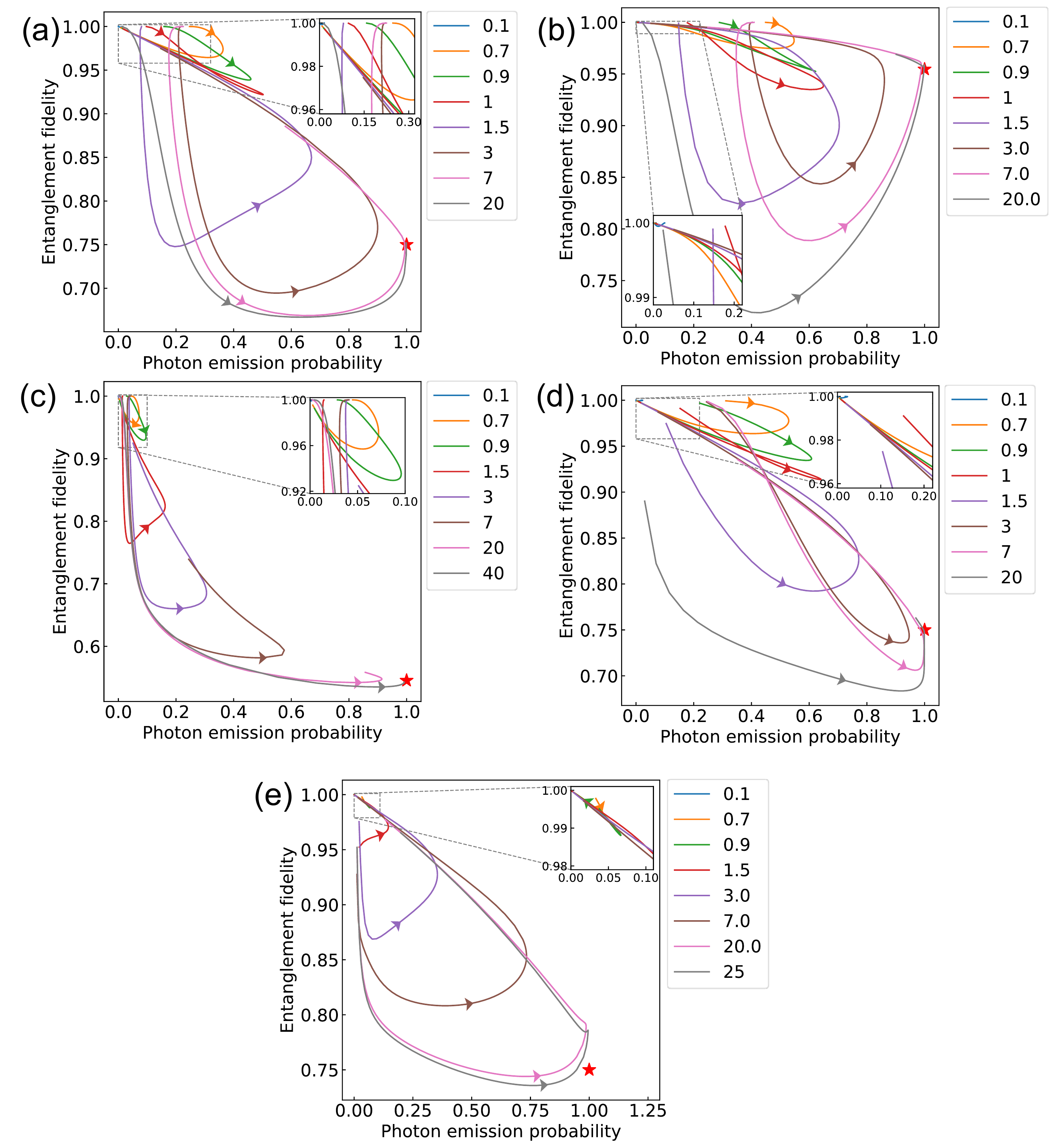}
    \caption{Relation between the entanglement fidelity $F$ and the photon emission probability $P_{\rm ex}$. In each trajectory, the pulse area $S$ is kept constant while the pulse width $\sigma$ is swept from 0.01 to 50.0. The pulse areas are color-coded as shown in the legends. 
    The red stars correspond to $F = 0.5+0.5C/(C+1)$ in Eq.~(\ref{eq:eq55}). (a)-(e) correspond to the parameter settings (a)-(e) in table~\ref{table:Tab1} respectively. The insets expand the areas in the gray rectangles to discern the overlapping trajectories.}
    \label{fig:fig4}
\end{figure*}

In this subsection, we numerically evaluate the rate-fidelity trade-off of the remote entanglement generation described in Sec.~I and II (see also Fig.~\ref{fig:fig1}). 
The fidelity of the generated entangled state is given in Eq.~(\ref{eq:eq4}). 
The entanglement generation rate without photon losses is given by
\begin{equation}
    \label{eq:eq33}
    P_{\rm ent} = \frac{1}{2}P_{\rm ex}^2,
\end{equation}

where $P_{\rm ex}$ is the photon emission probability from each ion-cavity system, given in Eq.~(\ref{eq:eq21}), and 1/2 stands for the successes rate of the standard optical Bell measurement employed in Fig.~\ref{fig:fig1} (see e.g., \cite{Duan2004}) for the entanglement swapping. 
Since we are interested in the ion-cavity dynamics and idealize the Bell measurement process,  
hereafter, we evaluate the trade-off between $F$ and $P_{\rm ex}$.
The parameters used in the simulation are given in Table~\ref{table:Tab1}.

Figure \ref{fig:fig4} shows the relation between $F$ and $P_{\rm ex}$ for the various parameter sets listed in Table~\ref{table:Tab1} when a Gaussian pulse in the following form is used for the pump:
\begin{equation}
    \label{eq:eq34}
    \Omega(t)=\Omega_0\exp(-\frac{(t-t_c)^2}{2\sigma^2}),
\end{equation}
where $\Omega_0$ is the maximum Rabi frequency, $t_c$ is the central time, and $\sigma$ is the standard deviation of the Gaussian function. 
The different trajectories in each subplot of Fig.~\ref{fig:fig4} correspond to different pulse areas $S$ whereas Figs.~\ref{fig:fig4}(a)-(e) correspond to the parameter sets (a)-(e) in Table~\ref{table:Tab1} respectively. Here, the pulse area is
\begin{equation}
    \label{eq:eq35}
    S=\frac{1}{\sqrt{2\pi}}\int \Omega (t) dt = \Omega_0\sigma.
\end{equation}
Within each trajectory, $\sigma$ changes from 0.01 to 50.0 while the pulse area is kept constant.
Note that due to the constant $S$, the maximum Rabi frequency $\Omega_0$ changes with $\sigma$ in accordance with Eq.~(\ref{eq:eq35}). 
The rotating-wave approximation is valid for these parameter regions since $\gamma_u$ is sufficiently lower than the transition frequency of atoms.

\begin{figure*}[htbp]
    \centering
    \includegraphics[width=13truecm]{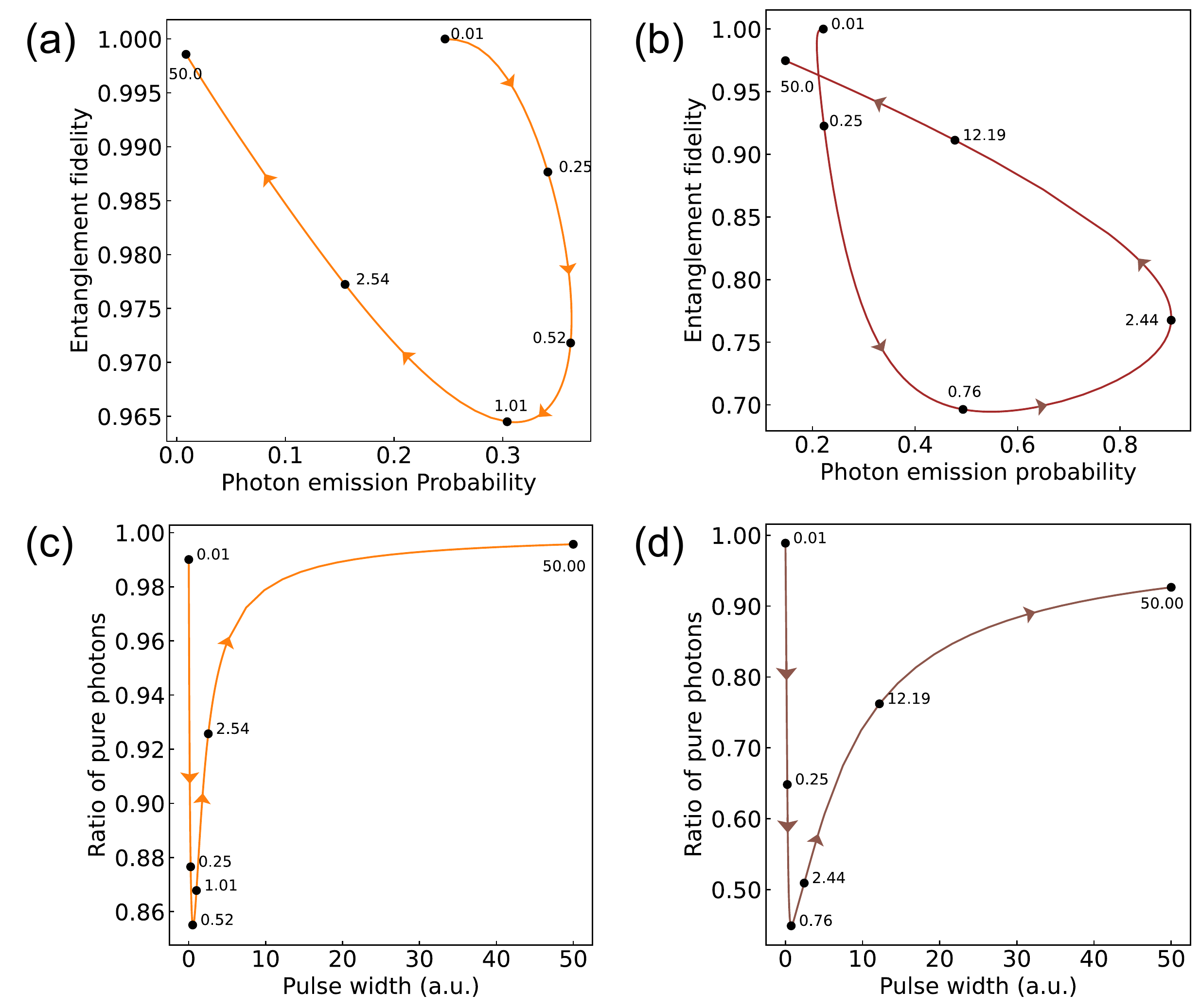}\\
    \caption{Here two trajectories ($S = 0.7$ and $3.0$) are selected from Fig.~\ref{fig:fig4}(a) and replotted in (a) and (b) respectively. The labels to the black dots on the trajectories indicate the corresponding pulse widths. (a) $F$ vs $P_{\rm ex}$ curve when $S = 0.7$. The numbers along with the black dots are the pulse widths. The trajectory of the plots from short to long pulses show a clockwise curve. (b) The $F$ vs $P_{\rm ex}$ curve when $S = 3.0$. The trajectory is counterclockwise from short to long pulses. 
    (c) The relationship between the pulse width and the ratio of a pure photon $P_{\rm pure}/P_{\rm ex}$ when $S=0.7$, where $P_{\rm pure}$ is defined in Eq.~(\ref{eq:eq36}) and Eq.~(\ref{eq:eq36-2}). The black dots correspond to the same pulse widths as in (a). 
    (d) The relationship between the pulse width and the ratio of a pure photon for $S=3.0$. The black dots correspond to the same pulse widths as in (b).}
    \label{fig:fig5}
\end{figure*}

All the curves in Fig.~4 show some trade-off between $F$ and $P_{\rm ex}$. To see how these trade-off curves (trajectories) behave with respect to the pulse width,
we pick up the trajectories with pulse areas of $0.7$ and $3.0$ in Fig.~\ref{fig:fig4}(a) and replot them in Fig.~\ref{fig:fig5}(a) and (b) respectively. 
The black dots are data points in the trajectories at particular pulse widths as annotated in the figures.
In Fig.~\ref{fig:fig5}(a), as the trajectory makes a loop, for the same $P_{\rm ex}$ there are two possible pulse widths $\sigma$ and the one with the smaller $\sigma$ exhibits a higher fidelity. 
In other words, as the pulse width increases, the trajectory makes a clockwise loop. 
In Fig.~\ref{fig:fig5}(b), on the other hand, the loop is in the counterclockwise direction with the increasing pulse width. Therefore the longer pulse (i.e. larger $\sigma$) achieves higher a fidelity for a given $P_{\rm ex}$ except for $P_{\rm ex} \lesssim 0.22$. 
These results indicate that depending on the pump pulse area, there is a qualitative difference in how the relationship between $F$ and $P_{\rm ex}$ evolves when changing the pump pulse width. 
Fig.~\ref{fig:fig5}(c) and (d) show the relative probability $P_{0}/P_{\rm ex}$ as a function of $\sigma$ where $P_0$ is the probability for the single photon generation that does not suffer from spontaneous emissions: 
\begin{equation}
    \label{eq:eq36}
    P_0=2\kappa\int\langle g1|\rho_0(t)|g1\rangle dt,
\end{equation}
where $\rho_0$ is the solution of the following differential equation,
\begin{equation}
    \label{eq:eq36-2}
    \frac{d}{dt}\rho_0(t)=-i(H_{\rm eff}\rho_0(t)-\rho_0(t)H_{\rm eff}^\dagger).
\end{equation}
As Eq.~(\ref{eq:eq36-2}) does not include the dissipative term $2\gamma \langle e0|\rho(t)| e0\rangle |u0\rangle\langle u0|$ the ratio $P_0/P_{\rm ex}$ represents the ratio of the single photon that is not affected by re-excitation. The black dots in Fig.~\ref{fig:fig5}(c) and (d) correspond to the same values of $\sigma$ as in Fig.~\ref{fig:fig5}(a) and (b). 
The trend of the fidelity and the ratio $P_0/P_{\rm ex}$ are well matched, which indicates suppression of the spontaneous decay increases the fidelity. The effect of the spontaneous decay is suppressed in the regime where the pulse width is either much shorter or much longer than the characteristic time of the decay ($\approx 1/\gamma_u =1$). 
In the regime of short pulse widths ($\sigma \ll 1$), even if the atom suffers from a spontaneous emission it is unlikely for the atom to be re-excited since the pump pulse is so short that after the spontaneous decay there is no appreciable intensity left in the pump pulse. 
Thus, the fidelity can be high at the expense of small $P_{\rm ex}$. 
This regime is often used in single photon emission experiments with atoms without a cavity~\cite{Higginbottom2016}. 
In the regime of long pulse widths ($\sigma \gg 1$), the following conditions of vSTIRAP are partially satisfied~\cite{Kuhn1999,AxelKuhn2002}.
\begin{align}
    \abs{\frac{2g\dot{\Omega}(t)}{(\Omega(t))^2+g^2}}&\ll\frac{1}{2}\abs{\Delta\pm\sqrt{g^2+(\Omega(t))^2+\Delta^2}}\label{eq:vSTIRAP1}\\
    g&\ll \max(\Omega(t))\label{eq:vSTIRAP2}\\
    \kappa,\gamma_u&\ll g\label{eq:vSTIRAP3}.
\end{align}
Here, Eq.~\ref{eq:vSTIRAP1} is the adiabatic condition of the pump pulse, Eq.~(\ref{eq:vSTIRAP2}) corresponds to perfect transfer from $\ket{u0}$ to $\ket{g1}$, and Eq.~(\ref{eq:vSTIRAP3}) is the strong coupling condition. Long pulse widths satisfy Eq.~(\ref{eq:vSTIRAP1}) and then suppress the spontaneous emission.

However, if the maximum Rabi frequency $\Omega_0$ is smaller than the coupling $g$, Eq~(\ref{eq:vSTIRAP2}) is not satisfied, and then $P_{\rm ex}$ remains small as shown in Fig.~\ref{fig:fig5}(a).
If the maximum Rabi frequency is large enough, the benefit of vSTIRAP becomes more apparent as is shown in Fig~\ref{fig:fig5}(b).
Eq.~(\ref{eq:vSTIRAP3}) is not satisfied with the parameters for Fig.~\ref{fig:fig5}, leading to the imperfection of vSTIRAP and degradation of the fidelity.

Let us discuss each subplot of Fig.~\ref{fig:fig4} in more details. 
In Fig.~\ref{fig:fig4}(a), the loops are clockwise with increasing $\sigma$ when the pulse area $S$ is less than one and counterclockwise when $S$ is larger than one. This distinction is ambiguous around $S = 1$. 
With a small pulse area ($S<1$), the shorter pulse width gives a higher fidelity at a given $P_{\rm ex}$ in each trajectory. On the other hand, with a large pulse area ($S>1$), the longer pulse width gives a higher fidelity in each trajectory. The overall fidelities with the small pulse areas ($S<1$) are higher than those with the larger pulse areas ($S>1$). However, the maximum $P_\mathrm{ex}$ with small pulse areas is limited below $\sim0.5$ due to the low pulse energies. Thus, the large pulse areas are necessary to obtain $P_\mathrm{ex}$ approaching unity. 

In Fig.~\ref{fig:fig4}(b), the coupling between the atom and the cavity is larger than that in Fig.~\ref{fig:fig4}(a). Therefore, the overall fidelity increases. The boundary for the clockwise and counterclockwise trajectories is between $S = 0.9$ and $S = 1$. The counterclockwise trajectories exhibit higher fidelities than the clockwise trajectories except for the part of the trajectories  of $S = 0.7$ and $S = 0.9$ as shown in Fig.~\ref{fig:fig4}(b). This is because the strong coupling condition facilitates vSTIRAP as in Eq.~(\ref{eq:vSTIRAP3}). Comparing Fig.~\ref{fig:fig4}(a) and (b), the area where the clockwise pulse is preferential decreases, due to the high cooperativity.

In Fig.~\ref{fig:fig4}(c), the coupling constant is smaller than that in Fig.~\ref{fig:fig4}(a). 
Note that $\kappa=\gamma_u=10$ is chosen here just to save the computational time. 
The same trend as in Fig.~\ref{fig:fig4}(a) is observed. The boundary between the clockwise and counterclockwise trajectories is at around $S = 1$.
The maximum $P_{\rm ex}$ that can be obtained with the clockwise trajectories is further limited in comparison with Fig.~\ref{fig:fig4}(a). 
This is because longer pulses with large pulse areas than in Fig.~\ref{fig:fig4}(a) are needed to obtain a high probability of photon emission due to the small cooperativity. As a result, the fidelities are lowered due to spontaneous emissions that occur in long pulses.

In Fig.~\ref{fig:fig4}(d), we consider a case where the decay rate of the cavity is dominantly large whereas the cooperativity remains the same as Fig.~\ref{fig:fig4}(a). The plots look similar to those in Fig.~\ref{fig:fig4}(a).
This indicates that as long as the cooperativity is the same, the behavior of the $F$-$P_{\rm ex}$ trade-off only has a weak dependence on the cavity decay rate. 
This is reasonable since in our model scattering of the cavity photons to the free space is not taken account and hence all the generated photons are outcoupled from the cavity and used for the entanglement swapping.

In Fig.~\ref{fig:fig4}(e), we consider the case where the decay rate of the atom is dominantly large. In this case, all the loops are counterclockwise. This is due to the excessively large value of the spontaneous emission rate. A clockwise loop could be obtained by using even smaller $S$, but this has not been done in this study because of the unreasonably long computing time that was required.

\begin{figure*}[htbp]
    \centering
    \includegraphics[width=13truecm]{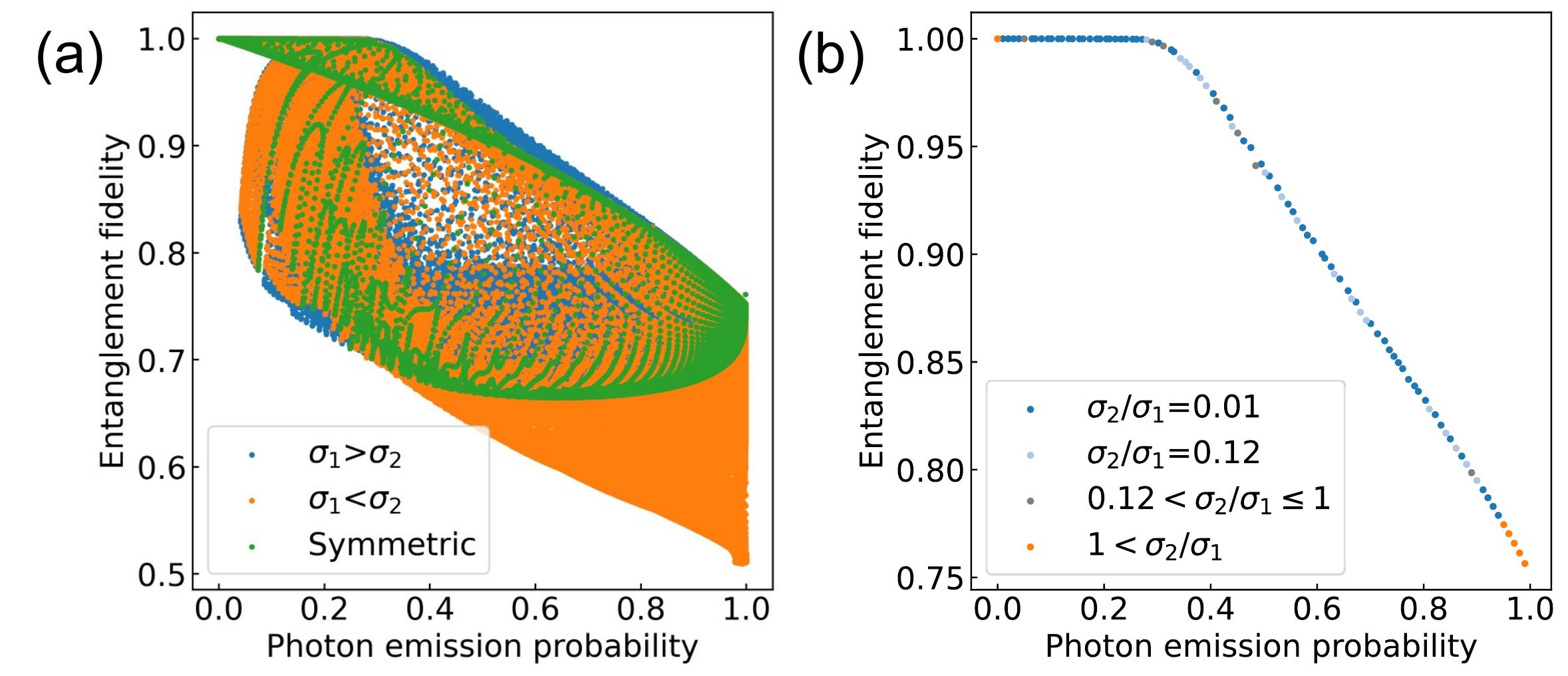}
    \caption{(a) Scatter plots of the $F-P_{\rm ex}$ relation with asymmetric as well as symmetric Gaussian pumping. Blue: asymmetric Gaussian pulses  $\sigma_1>\sigma_2$. Orange: asymmetric Gaussian pulses $\sigma_1<\sigma_2$. Green: symmetric Gaussian pulses. (b) The maximum fidelities extracted from (a). $P_{\rm ex}$ is binned into fixed intervals and at each interval the maximum fidelity is taken and plotted. }
    \label{fig:fig6}
\end{figure*}

Lastly we examine non-Gaussian pump pulses to explore the possibility of beating the trade-off imposed by the Gaussian pulses. Fig.~\ref{fig:fig6} shows the trade-off between photon emission probability and entanglement fidelity for various symmetric and asymmetric Gaussian pulses in the intermediate coupling regime ((a) in Table~\ref{table:Tab1}). The green points show the trade-off for the symmetric Gaussian pulses, which is a scatter plot equivalent to Fig.~\ref{fig:fig4}(a) but with a higher density. On the other hand, the blue and orange points show the trade-off for the asymmetric Gaussian pulses defined as
\begin{equation}
    \label{eq:eq37}
    \Omega(t)=\left\{
    \begin{array}{ll}
         \frac{2\Omega_0}{1+\sqrt{r}}\exp{-\frac{(t-t_c)^2}{2\sigma_1^2}}& t\leq t_c \\
         \frac{2\Omega_0}{1+\sqrt{r}}\exp{-\frac{(t-t_c)^2}{2\sigma_2^2}}& t\geq t_c 
    \end{array}
    \right.,
\end{equation}
where $r$ is the ratio of $\sigma_1$ and $\sigma_2$ while the pulse area is still given by Eq.~(\ref{eq:eq35}). The blue points represent the trade-off for $\sigma_1>\sigma_2$ and the orange points represent the trade-off for $\sigma_1 < \sigma_2$. The calculation method is as follows. First, we change $\sigma_2$ from $0.01\sigma_1$ to $\sigma_1$ while $\sigma_1$ is fixed and calculate $F$ and $P_{\rm ex}$. Next, we iterate the same calculation in the range of $\sigma_1=0.01$ to $20$ as shown in blue points in Fig.~\ref{fig:fig5}(a). Finally, we swap $\sigma_1$ for $\sigma_2$ and calculate the same in the previous procedures as shown in orange points in Fig.~\ref{fig:fig5}(a).  

As can be seen in Fig.~\ref{fig:fig6}(a), there is a region where asymmetric Gaussian pulses with $\sigma_1 > \sigma_2$ improves the fidelity over the Gaussian pulses for a range of $P_{\rm ex}$ (seen as blue dots above the boundary of the green dots). A few percent improvements from the Gaussian pulses in terms of the fidelity are observed in this region.
The Fig.~\ref{fig:fig6}(b) shows the maximum fidelity at a given $P_{\rm ex}$ in Fig.~\ref{fig:fig6}(a) where $P_\mathrm{ex}$ is binned into an interval of 0.01. It can be seen that most points are realized with $\sigma_2$ being significantly smaller than $\sigma_1$. In other words, the pump pulse should be turned off as soon as the transfer efficiency of vSTIRAP reaches the maximum.

\subsection{Analytical results}\label{subsection3.2}

In this subsection, we derive an approximate analytical expression for the fidelity Eq.~(\ref{eq:eq3}). 
Specifically, we clarify the relation between the fidelity and the cooperativity parameter in the limit of large $P_{\rm ex}$. 
The cooperativity $C$ is defined as
\begin{equation}
    \label{eq:eq38}
    C=\frac{g^2}{\kappa\gamma_u},
\end{equation}
which plays an important role in characterising atom-cavity systems.

The following analysis includes the off-resonant regimes,
\begin{align} 
    \Delta_p&\neq0\label{eq:eq39},\\
    \Delta_c&\neq0\label{eq:eq40}.
\end{align}
Taking an appropriate rotating frame, the effective Hamiltonian including these detunings is~\cite{Goto2019}
\begin{equation}
    \label{eq:eq41}
    H_\mathrm{eff}=
    \begin{bmatrix}
    \Delta_u & 0 & \Omega(t)\\
    0 & -i\kappa & g\\
    \Omega^*(t) & g & \Delta_e-i\gamma_u\\
\end{bmatrix},
\end{equation}
where $\Delta_u$ and $\Delta_e$ are the one and two photon detunings used in~\cite{Goto2019}. 
They are related to the detunings in the previous sections according to the relations $\Delta_c=\Delta_e$ and $\Delta_p=\Delta_e-\Delta_u$. 
Let us replace Eq.~(\ref{eq:eq26}) with Eq.~(\ref{eq:eq41}) in the master equation and the QRT. Taking the trace of the master equation Eq.~(\ref{eq:eq30}), we get 
\begin{eqnarray}
    \label{eq:eq42}        
    &&\Tr\left(\frac{d\rho_{\lambda}(t,t')}{dt'}\right) \nonumber\\
        &&=-i\Tr\left(H_\mathrm{eff}\rho_{\lambda}(t,t')-\rho_{\lambda}(t,t')H_\mathrm{eff}^\dagger\right),
\end{eqnarray}
and 
\begin{eqnarray}
 \label{eq:eq43}
        &&\frac{d}{dt'}\left(\abs{\lambda_1(t,t')}^2+\abs{\lambda_2(t,t')}^2+\abs{\lambda_3(t,t')}^2\right) \nonumber \\
        &&=-2\kappa \abs{\lambda_2(t,t')}^2-2\gamma_u\abs{\lambda_3(t,t')}^2.
\end{eqnarray}
Integrating the both sides of Eq.~(\ref{eq:eq43}) over $t'$ from $t$ to $\infty$ and $t$ from $0$ to $\infty$ reads
\begin{eqnarray}
    \label{eq:eq44}
        &&\int_0^\infty\abs{\lambda_1(t,t)}^2+\abs{\lambda_2(t,t)}^2+\abs{\lambda_3(t,t)}^2dt\nonumber\\
        &=&\int_0^\infty\abs{\lambda_1(t,\infty)}^2dt\nonumber\\&+&\int_0^\infty\int_t^\infty2\kappa\abs{\lambda_2(t,t')}^2+2\gamma_u\abs{\lambda_3(t,t')}^2dt'dt,
\end{eqnarray}
where we use $\lim_{t'\rightarrow\infty}\lambda_2(t,t')=\lim_{t'\rightarrow\infty}\lambda_3(t,t')=0$ since when $\kappa$ and $\gamma_u$ are finite, all the populations except $|u0\rangle$ and $|g0\rangle$ goes to zero. 
 
First, we consider the left-hand side of Eq.~(\ref{eq:eq44}). The term inside the integral is the population of the initial state for the Schr\"odinger equation Eq.~(\ref{eq:eq30}) and can be represented by the elements of the density matrix as,
\begin{eqnarray}
    \label{eq:eq45}
        &&\abs{\lambda_1(t,t)}^2+\abs{\lambda_2(t,t)}^2+\abs{\lambda_3(t,t)}^2\nonumber\\&=&\abs{\rho_{12}(t)}^2+\abs{\rho_{22}(t)}^2+\abs{\rho_{32}(t)}^2.
\end{eqnarray}
If the state is pure,
\begin{eqnarray}
    \label{eq:eq46}
            &&\abs{\rho_{12}(t)}^2+\abs{\rho_{22}(t)}^2+\abs{\rho_{32}(t)}^2\nonumber\\&=&\rho_{22}(t)(\rho_{11}(t)+\rho_{22}(t)+\rho_{33}(t)).
\end{eqnarray}
The decay rates $\kappa$ and $\gamma_u$ both contribute to the amplitude damping. When we focus on the system in the cavity, i.e. the upper-left $3\times3$ block matrix of Eq.~(\ref{eq:eq27}), $\kappa$ decreases the trace $\rho_{11}(t)+\rho_{22}(t)+\rho_{33}(t)$ and off-diagonal elements, but $\gamma_u$ does not decrease the trace. Taking these into account, we get the following inequality,
\begin{eqnarray}
    \label{eq:eq47}
        &&\int_0^\infty\abs{\rho_{12}(t)}^2+\abs{\rho_{22}(t)}^2+\abs{\rho_{32}(t)}^2dt\nonumber\\
        &\leq&\int_0^\infty\rho_{22}(t)(\rho_{11}(t)+\rho_{22}(t)+\rho_{33}(t))dt\nonumber\\
        &=&\int_0^\infty\rho_{22}(t)(1-\rho_{44}(t))dt,
\end{eqnarray}
where we use $\Tr(\rho)=1$ to get the third line. 
$\rho_{44}(t)$ is the probability of emitting a photon by time $t$ and we have
\begin{align}
    \rho_{44}(t)&=2\kappa\int_0^t\rho_{22}(t)dt, \label{eq:eq48}\\
    P_\mathrm{ex}&=\lim_{t\rightarrow \infty}\rho_{44}(t)=2\kappa\int_0^\infty\rho_{22}(t)dt\label{eq:eq49}.
\end{align}
Thus, the integration in the third line of Eq.~(\ref{eq:eq47}) is 
\begin{eqnarray}
    \label{eq:eq50}
        &&\int_0^\infty\rho_{22}(t)(1-\rho_{44}(t))dt \nonumber\\
        &&=\int_0^\infty\frac{1}{2\kappa}\frac{d\rho_{44}(t)}{dt}(1-\rho_{44}(t))dt \nonumber\\
        &&=\frac{P_\mathrm{ex}}{2\kappa}-\frac{P_\mathrm{ex}^2}{4\kappa}.
    \end{eqnarray}

Regarding the right-hand side of Eq.~(\ref{eq:eq44}), the first term $\lambda_1(t,\infty)$ is difficult to break down since it strongly depends on the waveform of the pump pulse, although in the limit of $P_\mathrm{ex}\rightarrow1$, it goes to zero. This is because it represents the remaining population in the cavity. The second term is similar to the waveform correlation $\langle J\rangle$ in Eq.~(\ref{eq:eq18}) except for the integration range. However, we have the following relation: 
\begin{eqnarray}
    \label{eq:eq51}
        &&2\kappa\int_0^\infty\int_t^\infty\abs{\lambda_2(t,t')}^2dt'dt\nonumber\\
        &=&\kappa\int_0^\infty\int_0^\infty\abs{\lambda_2(t,t')}^2dt'dt=\frac{P_\mathrm{ex}^2}{4\kappa}\langle J\rangle, 
\end{eqnarray}
where the first equality is due to the the symmetry of $\langle a^\dagger(t)a(t')\rangle$ with respect to $t$ and $t'$ and the second equality is derived from Eq.~(\ref{eq:eq18}) and Eq.~(\ref{eq:eq32}).
For the third term, we use the differential equation for $\lambda_2(t,t')$ in Eq.~(\ref{eq:eq30}): 
\begin{equation}
    \label{eq:eq52}
    \frac{d}{dt'}\lambda_2(t,t')=-\kappa\lambda_2(t,t')-ig\lambda_3(t,t').
\end{equation}
Let us substitute this equation to the third term and erase $\lambda_3(t,t')$. Then we have
\begin{widetext}
    \begin{eqnarray}
    \label{eq:eq53}
        &&\int_0^\infty\int_t^\infty\frac{2\gamma_u}{g^2}\abs{\frac{d}{dt'}\lambda_2(t,t')+\kappa\lambda_2(t,t')}^2dt'dt\nonumber\\
        &=&\int_0^\infty\int_t^\infty\frac{2}{\kappa C}\abs{\frac{d}{dt'}\lambda_2(t,t')}^2+\frac{4}{C}\Re\left(\lambda_2^*(t,t')\frac{d}{dt'}\lambda_2(t,t')\right)+\frac{2\kappa}{C}\abs{\lambda_2(t,t')}^2dt'dt\nonumber\\
        &=&\int_0^\infty\int_t^\infty\frac{2}{\kappa C}\abs{\frac{d}{dt'}\lambda_2(t,t')}^2dt'dt-\int_0^\infty\frac{2}{C}\abs{\lambda_2(t,t)}^2dt+\frac{P_\mathrm{ex}^2}{4\kappa C}\langle J\rangle,
    \end{eqnarray}
where $C$ is the cooperativity defined in Eq.~(\ref{eq:eq38}).
Finally, substituting these equations into Eq.~(\ref{eq:eq44}) and after some algebra, we have the following inequality
\begin{eqnarray}
    \label{eq:eq54}
        \langle J\rangle &\leq& \frac{2C}{C+1}(\frac{1}{P_\mathrm{ex}}-\frac{1}{2})+\int_0^\infty\frac{8\kappa}{P_\mathrm{ex}^2(C+1)}\abs{\lambda_2(t,t)}^2dt\nonumber\\
        &&-\int_0^\infty\int_t^\infty\frac{8}{(C+1)P_\mathrm{ex}^2}\abs{\frac{d}{dt'}\lambda_2(t,t')}^2dt'dt-\frac{4\kappa C}{P_\mathrm{ex}^2(C+1)}\int_0^\infty\abs{\lambda_1(t,\infty)}^2dt. 
\end{eqnarray}
Especially, when $P_\mathrm{ex}\rightarrow1$, we get 
\begin{equation}
    \label{eq:eq55}
    \langle J\rangle \leq \frac{C}{C+1}+\int_0^\infty\frac{8\kappa}{C+1}\abs{\lambda_2(t,t)}^2dt
        -\int_0^\infty\int_t^\infty\frac{8}{(C+1)}\abs{\frac{d}{dt'}\lambda_2(t,t')}^2dt'dt
    \end{equation}
\end{widetext}
The first-term of the right-hand side is independent of the waveform of the driving pulse. 
Moreover, we numerically find that the contribution of the second and the third terms are relatively small. 
That is, Eq.~(\ref{eq:eq55}) can be approximated to 
\begin{equation}
    \label{eq:eq56}
    \langle J\rangle \lesssim \frac{C}{C+1}. 
\end{equation}
Combining it with Eq.~(\ref{eq:eq4}), we have 
\begin{equation}
    \label{eq:eq57}
    F \lesssim \frac{1}{2}\left( 1+ \frac{C}{C+1} \right),  
\end{equation}
in the limit of $P_\mathrm{ex}\rightarrow1$. 
In Figs.~\ref{fig:fig4}(a)-(e), we plot the right hand side of Eq.~(\ref{eq:eq57}) at $P_{\rm ex} = 1$ as red stars. 
All these plots except for Fig.~\ref{fig:fig4}(e) show a good agreement with the numerical simulation. 
It means that the inequality in Eq.~(\ref{eq:eq57}) is almost saturated by using Gaussian pump pulses and the maximum fidelity at this point can be simply characterized by $C$. In Fig.~\ref{fig:fig4}(e), however, the upper bound is 0.03 smaller than the numerical results of fidelity since the second and third terms in Eq.~(\ref{eq:eq55}) are no longer negligible.
It is interesting to note that Ref.~\cite{Goto2019} pointed that the expression $nC/(nC+m)$ ($n,m\in\mathbb{N}$) upper bounds various probabilities in an atom-cavity system such as the emitting photon, gate operations~\cite{Goto2008,Goto2010}, and photon absorption~\cite{Gorshkov2007,Dilley2012}. 
Our result adds $\langle J\rangle$ as a new example of such upper bounds. 

\subsection{Multi-partite entanglement swapping and the waveform of the photons}\label{subsection3.3}
\begin{figure}[htbp]
    \centering
    \includegraphics[width=8.6truecm]{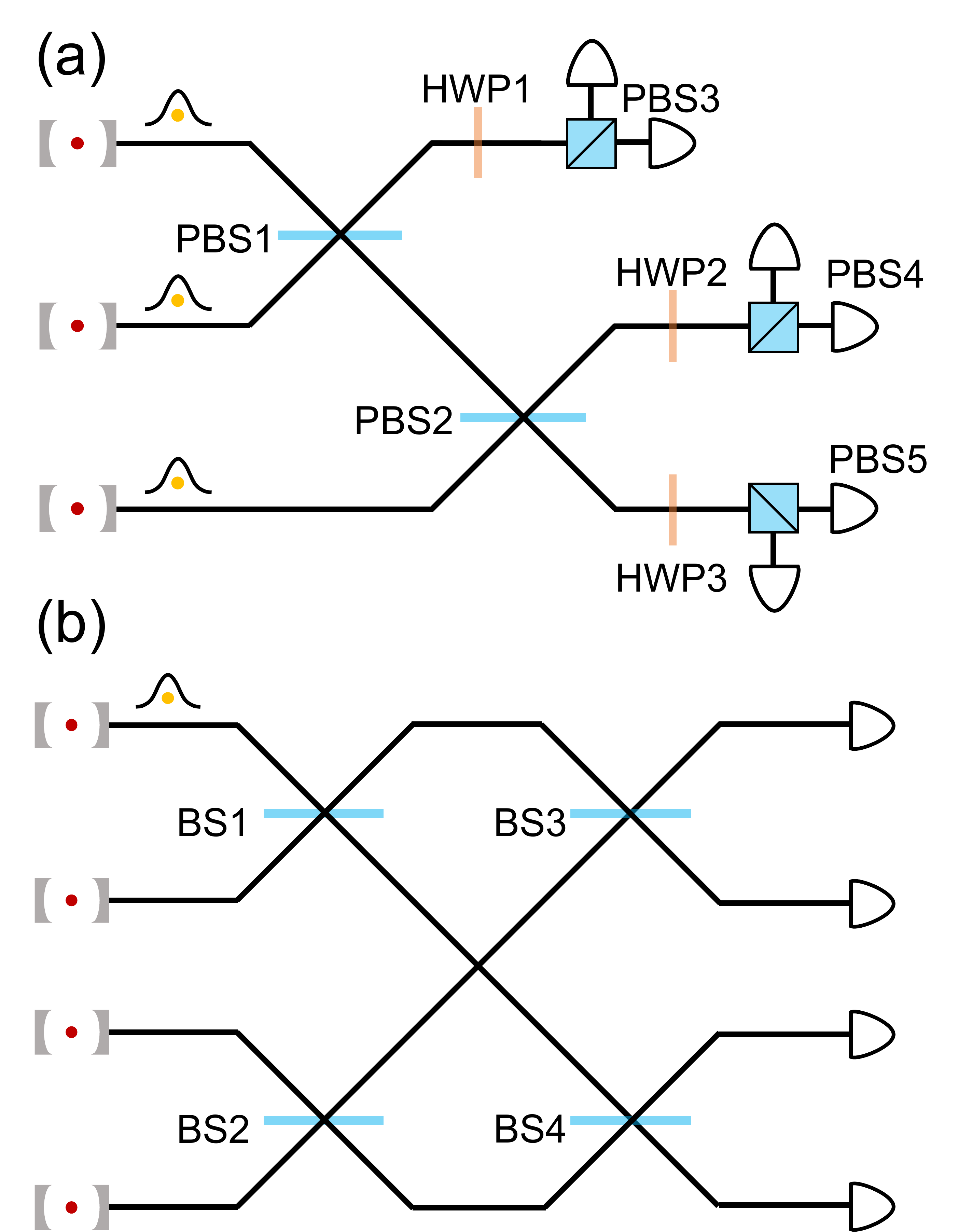}
    \caption{(a) Interferometer to generate GHZ state entanglement~\cite{Pan1998}. (b) Interferometer to generate a W-state~\cite{Lougovski2009,Roga2023}}
    \label{fig:Fig2}
\end{figure}
In this subsection, we extend the above scenario to multipartite entanglement sharing among more than two atom-cavity systems. 
Fig.~\ref{fig:Fig2} illustrates some examples of such scenarios including the GHZ state generation~\cite{Pan1998} and the W state generation~\cite{Lougovski2009,Roga2023}. 
Here we consider a general case of interconnecting $N$ atom-cavity systems. 
Then the state before photon detectors (i.e. after beam splitters) is described as 
\begin{equation}
    \label{eq:eq7}\ket{\Psi}_\mathrm{pre}^\mathrm{ideal}=\sum_{m=1}^M c_m \bigotimes_{n=1}^N (a_n^\dagger)^{s_{mn}}(\sigma_n^\dagger)^{t_{mn}}\ket{0}_p\ket{\downarrow}_a,
\end{equation}
where $N$ is the number of modes, $M$ is the number of terms in the superposition, $c_m$ is the probability amplitude of the $m$-th term, $a_n^\dagger$ is the photonic creation operator for the mode $n$, $\sigma_n^\dagger$ is the atomic creation operator for the mode $n$, and $s_{mn}=0,1$ and $t_{mn}=0,1$. 
The subscripts $p$ and $a$ denote the photon and atom subsystem respectively.
As in the previous subsection, we introduce the temporal waveform of the photons: 
\begin{eqnarray}
    \ket{\Psi}_\mathrm{pre}^\mathrm{real}&=&
    \sum_{m=1}^M c_m \bigotimes_{n=1}^N \left(A_{mn}^\dagger\right)^{s_{mn}}(\sigma_n^\dagger)^{t_{mn}}\ket{0}_p\ket{\downarrow}_a\label{eq:eq8},\\
    A_{mn}^\dagger&=&\int \phi_{mn}^*(t)a_n^\dagger(t) dt\label{eq:eq9},
\end{eqnarray}
where $\phi_{mn}(t)$ is the waveform of the photon. 
The state after the photo-detection is
\begin{equation}
    \label{eq:eq10}
    \ket{\Psi}_{\rm post}^{\rm real}=\sum_{m=\{k_1,...,k_K\}}c_m\prod_{n=\{l_1...l_L\}}\phi_{mn}^*(t_n)\ket{\psi_m}_a,
\end{equation}
where $\{k_1,...k_K\}$ denotes the terms remaining after the measurement, $\{l_1,...,l_L\}$ are the modes in which photons are detected, $t_n$ is the times when the photons are detected, and $\ket{\psi_m}_a=\otimes_{n=1}^N (\sigma_n^\dagger)^{t_{mn}}\ket{\downarrow}$. 
The details can be found in appendix~\ref{AppendixC.2}.
The fidelity of the atomic entanglement $\Bar{\mathcal{F}}$ is given as a time-averaged inner product between the target state 
\begin{eqnarray}
    \label{eq:eq11}
    \ket{\psi}_{\rm post}^{\rm ideal}=\sum_{m=\{k_1,...,k_K\}} c_m \ket{\psi_m}_a
\end{eqnarray}
and $\ket{\psi}_{\rm post}^{\rm real}$, as follows,
\begin{widetext}
    \begin{eqnarray}
    \label{eq:eq12}
        \Bar{\mathcal{F}}&=&\int\cdots\int dt_1\cdots dt_{l_L}\abs{{}_\mathrm{post}^\mathrm{ideal}\braket{\psi}{\psi}_\mathrm{post}^\mathrm{real}}^2\nonumber\\
        &=&\int\cdots\int dt_1\cdots dt_{l_L}\abs{\sum_{m=\{k_1,...,k_K\}}c_m^*\bra{\psi_m}_a\sum_{m'=\{k_1,...,k_K\}}c_{m'}\prod_{n'=\{l_1...l_L\}}\phi_{m'n'}^*(t_n')\ket{\psi_m'}_a}^2\nonumber\\
        &=&\int\cdots\int dt_1\cdots dt_{l_L}\left|\sum_{m, m^{\prime}=\left\{k_1, \ldots, k_K\right\}} I_{m m^{\prime}} \prod_{n^{\prime}=\left\{l_1. \ldots, l_L\right\}} \phi_{m^{\prime} n^{\prime}}^*\left(t_{n^{\prime}}\right)\right|^2\nonumber\\
        &=&\sum_{m, m^{\prime}, p, p^{\prime}=\left\{k_1, \ldots, k_K\right\}} I_{m m^{\prime}}^* I_{p p^{\prime}} \bigotimes_{n=\left\{l_1, \ldots, l_L\right\}} \int \phi_{m^{\prime} n}\left(t_n\right) \phi_{p^{\prime} n}^*\left(t_n\right) d t_n.
\end{eqnarray}
\end{widetext}
Note that the inner product $I_{mm^\prime}=c_m^*c_{m^\prime}\braket{\psi_m}{\psi_{m\prime}}_a$ is time-independent. Then, the fidelity only depends on the two-point correlation functions of the waveforms for each pair of the photons despite of the multi-photon interference. 
As a result, even in multi-partite entanglement generation scenarios, the fidelity of the generated state can always be evaluated by computing the two-photon correlation functions alone (in other words, higher order correlation functions are not needed). This means that the results of the previous sections are also relevant without modification in the case of multipartite entanglement generation.

\section{conclusion}\label{section4}
In this paper, we study the entanglement generation of two distant atom-cavity systems via photonic interconnects under the influence of spontaneous emission followed by re-excitation of the atoms. 
We demonstrated the quantitative relationship between the entanglement fidelity $F$, and the photon emission rate $P_{\rm ex}$ where the latter directly corresponds to the entanglement generation rate. We study the trade-off between $F$ and $P_{\rm ex}$ on the parameters of the Gaussian-shaped pump pulse.

We have found that there are qualitatively distinct regimes depending on the value of the pulse area, characterized by clockwise and counterclockwise trajectories respectively in the $F - P_{\rm ex}$ plane as the pulse width is changed. 
In the former when the pulse area is relatively small, the shorter pulse width gives a higher fidelity for a given $P_{\rm ex}$, where the pulse width is much smaller than the characteristic time of the decay ($\sigma\ll1/\gamma_u$). In the latter with relatively large pulse areas, the longer pulse width ($\sigma\gg1/\gamma_u$) gives a higher fidelity. 
In other words, with small pulse areas non-adiabatic pumping is more preferential than adiabatic pumping and this relationship is reversed when large pulse areas are used.
For the weak and intermediate regimes of coupling, using smaller pulse areas achieves higher fidelities than larger pulse areas at the expense of achievable $P_{\rm ex}$. However, as the atom-cavity coupling increases this difference in the fidelity becomes negligible and using large pulse areas becomes more beneficial as it achieves high $F$ and high $P_{\rm ex}$ at the same time as the effectiveness of vSTIRAP improves.

In addition, we show that asymmetric Gaussian pump pulses improve the trade-off with a few percent increase in the fidelity. Thus, numerical optimization may allow further improvements of the trade-off by exploring more general waveforms of pump pulses.

These results provide a guideline as to how to choose the pump pulse waveform for the photonic interconnects of the atom-cavity systems.

We also analytically derive an upper bound for the entanglement fidelity in the limit of unit photon emission probability. Combining it with numerical evidences, we give a simple expression of the approximated maximum fidelity in this limit, which is solely a function of the cooperativity of the atom-cavity system.
Interestingly, the expression obtained here is a class of $nC/(nC+m)$ which upper bounds various probabilities in the atom-cavity systems. 

In addition, we extend our bipartite entanglement generation model to multi-partite entanglement including the GHZ and W state generation. 
We consider a general Bell-measurement-like photonic interconnects and show that for any configuration and any number of parties in this setting, the fidelity of the generated entanglement can be evaluated by the two-photon correlation functions.
Therefore our results on the pump pulses are also directly applicable in the multipartite scenarios.
Our findings will drastically simplify the analyses of the future quantum network based on the Bell-measurement-like photonic interconnects. 

\begin{acknowledgments}
This work is supported by JST Moonshot R\&D Grant Nos. JPMJMS2066 and JPMJMS2063.
\end{acknowledgments}

\appendix

\section{The Bell state fidelity in entanglement swapping, considering the waveform of a photon}\label{AppendixA}
The starting point is the wavefunction of the input state in Fig.~\ref{fig:fig1} which is given in Eq.~(\ref{eq:eq1}). The Beam splitter transforms the creation and annihilation operators as follows:
\begin{eqnarray}
    \label{eq:eq.A.1}
    \begin{bmatrix}
    a_1^\mu\\
    a_2^\mu
    \end{bmatrix}=\frac{1}{\sqrt{2}}
    \begin{bmatrix}
    1&-1\\
    1&1
    \end{bmatrix}
    \begin{bmatrix}
    a_3^\mu\\
    a_4^\mu
    \end{bmatrix}.
\end{eqnarray}
Using Eq.~(\ref{eq:eq2}) and Eq.~(\ref{eq:eq.A.1}) let us define the time-dependent annihilation operator at output modes as follows,
\begin{equation}
    \label{eq:eq.A.2}
    A_{(i,j)}^\mu=\int dt \phi_{i}^{\mu*}(t)a_j^\mu(t),
\end{equation}
where $i=1,2$ denotes the input modes, and $j=3,4$ denotes the output modes.
Then, the output state reads
\begin{widetext}
\begin{eqnarray}
    \label{eq:eq.A.3}
        \ket{\Psi_\mathrm{out}}&=&\frac{1}{4}\left[\ket{\uparrow\uparrow}\left(A_{(1,3)}^{H\dagger}A_{(2,3)}^{H\dagger}+A_{(1,3)}^{H\dagger}A_{(2,4)}^{H\dagger}-A_{(1,4)}^{H\dagger}A_{(2,3)}^{H\dagger}-A_{(1,4)}^{H\dagger}A_{(2,4)}^{H\dagger}\right)\right.\nonumber\\
        &&+\ket{\downarrow\downarrow}\left(A_{(1,3)}^{V\dagger}A_{(2,3)}^{V\dagger}+A_{(1,3)}^{V\dagger}A_{(2,4)}^{V\dagger}-A_{(1,4)}^{V\dagger}A_{(2,3)}^{V\dagger}-A_{(1,4)}^{V\dagger}A_{(2,4)}^{V\dagger}\right)\nonumber\\
        &&+\ket{\uparrow\downarrow}\left(A_{(1,3)}^{H\dagger}A_{(2,3)}^{V\dagger}+A_{(1,3)}^{H\dagger}A_{(2,4)}^{V\dagger}-A_{(1,4)}^{H\dagger}A_{(2,3)}^{V\dagger}-A_{(1,4)}^{H\dagger}A_{(2,4)}^{V\dagger}\right)\nonumber\\
        &&+\left.\ket{\downarrow\uparrow}\left(A_{(1,3)}^{V\dagger}A_{(2,3)}^{H\dagger}+A_{(1,3)}^{V\dagger}A_{(2,4)}^{H\dagger}-A_{(1,4)}^{V\dagger}A_{(2,3)}^{H\dagger}-A_{(1,4)}^{V\dagger}A_{(2,4)}^{H\dagger}\right)\right]\ket{0}.
\end{eqnarray}
There are four sets of detection, which herald the generation of entanglement, $a_3^H a_3^V$, $a_3^H a_4^V$, $a_3^V a_4^H$ and $a_4^H a_4^V$. For example, let us consider $a_3^H a_3^V$. The post-measurement state is
\begin{eqnarray}
    \label{eq:eq.A.4}
        \ket{\psi_\mathrm{atom}(t)}&=&\frac{1}{\mathcal{N}}\bra{0}a_3^H(t) a_3^V(t')\ket{\Psi_{out}}\nonumber\\
        &=&\frac{1}{4\mathcal{N}}\bra{0}a_3^H(t) a_3^V(t')\left(A_{(1,3)}^{H\dagger}A_{(2,3)}^{V\dagger}\ket{\uparrow\downarrow}+A_{(1,3)}^{V\dagger}A_{(2,3)}^{H\dagger}\ket{\downarrow\uparrow}\right)\ket{0}\nonumber\\
        &=&\frac{1}{4\mathcal{N}}\bra{0}\iint dsds'\phi_1^{H*}(s)\phi_2^{V*}(s')a_3^H(t) a_3^V(t')a_3^{H\dagger}(s)a_3^{V\dagger}(s')\ket{\uparrow\downarrow}\nonumber\\
        &&\hspace{18pt}+\iint dsds'\phi_1^{V*}(s)\phi_2^{H*}(s')a_3^H(t) a_3^V(t')a_3^{V\dagger}(s)a_3^{H\dagger}(s')\ket{\downarrow\uparrow}\ket{0}\nonumber\\
        &=&\frac{1}{4\mathcal{N}}\left(\phi_1^{H*}(t)\phi_2^{V*}(t')\ket{\uparrow\downarrow}+\phi_1^{V*}(t')\phi_2^{H*}(t)\ket{\downarrow\uparrow}\right),
\end{eqnarray}
\end{widetext}
where $\mathcal{N}=\frac{1}{4}\sqrt{\abs{\phi_1^{H}(t)}^2\abs{\phi_2^{V}(t')}^2+\abs{\psi_1^{V}(t')}^2\abs{\phi_2^{H}(t)}^2}$ is the normalization factor, and in the third line we use 
\begin{equation}
    \label{eq:eq.A.5}
    \left[a_i^\mu(t),a_i^\mu(s)\right]=\delta(t-s).
\end{equation}
$\ket{\psi_\mathrm{atom}(t)}$ is the time-dependent wavefunction. In a real experiment it is often not possible or not practical (e.g. due to low event rates) to distinguish photon arrival times at the detectors with narrow time windows. In that case, the state is described by the following time-averaged density operator
\begin{eqnarray}
    \label{eq:eq.A.6}
        \rho_\mathrm{atom}&=&\iint dtdt'\mathcal{N}^2\ketbra{\psi_{\mathrm{atom}}}{\psi_\mathrm{atom}}\nonumber\\
        &=&\frac{1}{16}\left(\ketbra{\uparrow\downarrow}{\uparrow\downarrow}+\ketbra{\downarrow\uparrow}{\downarrow\uparrow}\right.\nonumber\\
        &&\left.+J\ketbra{\uparrow\downarrow}{\downarrow\uparrow}+J^*\ketbra{\downarrow\uparrow}{\uparrow\downarrow}\right),
\end{eqnarray}
where $J$ is the same in Eq.~\ref{eq:eq5}. Finally, the fidelity between $\rho_{\mathrm{atom}}$ and $\ket{\Psi_\mathrm{Bell}^+}=\frac{1}{\sqrt{2}}(\ket{\uparrow\downarrow}+\ket{\downarrow\uparrow})$ is 
\begin{equation}
    \label{eq:eq.A.7}
    F=\frac{\bra{\Psi_\mathrm{Bell}^+}\rho_\mathrm{atom}\ket{\Psi_\mathrm{Bell}^+}}{\Tr[\rho_{\mathrm{atom}}]}=\frac{1}{2}(1+\Re{J}),
\end{equation}
as in Eq.~(\ref{eq:eq4}).

\section{Multi-partite entanglement swapping and waveform of photons}\label{AppendixC}
\subsection{Single-mode case (ideal case)}\label{AppendixC.1}
Let us consider a general situation in which single photons from multiple identical atomic sources go
through a network of beam splitters and are detected by photon detectors
heralding the generation of an entangled atomic state. The quantum state of the system just before
the photon detection can be written as follows:
\begin{equation}
    \label{eq:eq.B.1}
    \ket{\psi}_\mathrm{pre}^\mathrm{ideal}=\sum_{m=1}^M c_m \bigotimes_{n=1}^N (a_n^\dagger)^{s_{mn}}(\sigma_n^\dagger)^{t_{mn}}\ket{0}_p\ket{\downarrow}_a.
\end{equation}
Here $s_{mn}, t_{mn}=0$ or $1$, and $a_n^\dagger$ and $\sigma_n^\dagger$ are the creation operators for the photonic and atomic
degrees of freedom respectively. By choosing an appropriate combination of values for
$s_{mn}, t_{mn}\ (n=1,...,N)$, the term in Eq.~(\ref{eq:eq.B.1}) can represent an arbitrary state with at most one atomic
and photonic excitation per mode. Even though the total number of the atomic and photonic modes
are assumed to be the same ($=N$) in Eq.~(\ref{eq:eq.B.1}), this does not reduce the generality of the state: in case that the
number of the atomic modes $N_a$ is larger than that of the photonic modes $N_p$, one can always
consider additional $(N_a-N_p)$ virtual photonic modes that always remain in vacuum such that the
nominal number of the atomic and photonic modes coincide. A similar discussion applies when $N_p>N_a$.
Let us assume that single photons are detected in different output modes. The state after the detection is,
\begin{equation}
    \label{eq:eq.B.2}
    \ket{\psi}_\mathrm{post}^\mathrm{ideal}=\bra{0}_p a_{l_1}...a_{l_L}\ket{\psi}_\mathrm{pre}^\mathrm{ideal}.
\end{equation}
Among all $M$ terms in Eq.~(\ref{eq:eq.B.2}), only the terms that have the exact combination of the photon creation
operators $a_{l_1}^\dagger...a_{l_L}^\dagger$ can cancel the annihilation operators associated with the photon detection and
remain non-zero. Let us assume that there are $K$ such terms and we index them with $k_1,...,k_K$.
\begin{eqnarray}
    \label{eq:eq.B.3}
        \ket{\psi}_\mathrm{post}^\mathrm{ideal}&=&\sum_{m=\{k_1,...,k_K\}}c_m\bigotimes_{n=\{l_1...l_L\}}(\sigma_n^\dagger)^{t_{mn}}\ket{0}_a\nonumber\\
        &=&\sum_{m=\{k_1,...,k_K\}}c_m\ket{\psi_m}_a.
\end{eqnarray}
Here $\ket{\psi}_a=\otimes_{n=\{l_1...l_L\}}(\sigma_n^\dagger)^{t_{mn}}\ket{\downarrow}_a$.

\subsection{Including temporal modes (real case)}\label{AppendixC.2}
Taking the temporal modes of the photons into account can be done by replacing $a_n^\dagger$ with
$\int \phi_{mn}^*(t)a_n^\dagger(t) dt$ in Eq.~(\ref{eq:eq.B.1}). The state of the system prior to the photon detection becomes,
\begin{eqnarray}
    \label{eq:eq.B.4}
        &&\ket{\psi}_\mathrm{pre}^\mathrm{real}=
    \sum_{m=1}^M c_m \bigotimes_{n=1}^N \left(A_{mn}^\dagger\right)^{s_{mn}}(\sigma_n^\dagger)^{t_{mn}}\ket{0}_p\ket{\downarrow}_a,\nonumber\\
    &&\mathrm{where}\quad A_{mn}^\dagger=\int \phi_{mn}^*(t)a_n^\dagger(t) dt.
\end{eqnarray}
We assume photon detection happens in the $l_1,...,l_L$ modes at times $t_{l_1},...,t_{l_2}$ respectively. The post
detection state can be obtained by applying $\bra{0}_p a_{l_1}(t_{l_1})...a_{l_L}(t_{l_L})$ to Eq.~(\ref{eq:eq.B.4}),
\begin{equation}
    \label{eq:eq.B.5}
    \ket{\psi}_\mathrm{post}^\mathrm{real}=\bra{0}_p a_{l_1}(t_{l_1})...a_{l_L}(t_{l_L})\ket{\psi}_\mathrm{pre}^\mathrm{real}.
\end{equation}
As in the ideal case, only the terms that have $a_{l_1}^\dagger(t)...a_{l_L}^\dagger$ remain
non-zero. We assume there are $K$ such terms $k_1,...,k_K$. Due to the
commutation relation $[a_n(t),a_n^\dagger(s)=\delta(t-s)]$,
\begin{equation}
    \label{eq:eq.B.6}
    a_l(t_l)\left(\int \phi_{ml}^*(t)a_l^\dagger(t)dt\right)\ket{0}_p=\phi_{mn}^*(t_l)\ket{0}_p
\end{equation}
Therefore, commuting $a_{l_1}^\dagger(t)...a_{l_L}^\dagger(t)$ and $a_{l_1}(t)...a_{l_L}(t)$ leaves out a factor $\phi_{ml_1}^*(t_{l_1})...\phi_{ml_L}^*(t_{l_L})$
\begin{eqnarray}
    \label{eq:eq.B.7}
        &&\ket{\psi}_\mathrm{post}^\mathrm{real}\nonumber=\sum_{m=\{k_1,...,k_K\}}c_m\bigotimes_{n=\{l_1...l_L\}}
            \phi_{mn}^*(t_n)(\sigma_n^\dagger)^{t_{mn}}\ket{\downarrow}_a\nonumber\\
        &&=\sum_{m=\{k_1,...,k_K\}}c_m\prod_{n=\{l_1...l_L\}}\phi_{mn}^*(t_n)\ket{\psi_m}_a.
\end{eqnarray}
\subsection{Fidelity}\label{AppendixC.3}
We calculate the fidelity between $\ket{\psi}_\mathrm{post}^\mathrm{ideal}$ and $\ket{\psi}_\mathrm{post}^\mathrm{real}$. From Eq.~(\ref{eq:eq.B.3}) and Eq.~(\ref{eq:eq.B.7}) we have:
\begin{widetext}
    \begin{eqnarray}
        \mathcal{F}(t_1,...,t_{l_L})&=&\abs{{}_\mathrm{post}^\mathrm{ideal}\braket{\psi}{\psi}_\mathrm{post}^\mathrm{real}}^2\nonumber\\
        &=&\abs{\sum_{m=\{k_1,...,k_K\}}c_m^*\bra{\psi_m}_a\sum_{m'=\{k_1,...,k_K\}}c_m'\prod_{n'=\{l_1...l_L\}}\phi_{m'n'}^*(t_n')\ket{\psi_m'}_a}^2\nonumber\\
        &=&\left|\sum_{m, m^{\prime}=\left\{k_1, \ldots ,k_K\right\}} c_m^* c_{m^{\prime}} \prod_{n^{\prime}=\left\{l_1, \ldots, l_L\right\}} \phi_{m^{\prime} n^{\prime}}^*\left(t_{n^{\prime}}\right)_a\left\langle\psi_m \mid \psi_{m^{\prime}}\right\rangle_a\right|^2\nonumber\\
        &=&\left|\sum_{m, m^{\prime}=\left\{k_1, \ldots, k_K\right\}} I_{m m^{\prime}} \prod_{n^{\prime}=\left\{l_1. \ldots, l_L\right\}} \phi_{m^{\prime} n^{\prime}}^*\left(t_{n^{\prime}}\right)\right|^2.
\end{eqnarray}
Here we used $I_{m m^{\prime}}=c_m^* c_{m^{\prime} a}\left\langle\psi_m \mid \psi_{m^{\prime}}\right\rangle_a$. Expanding the last expression, we get
\begin{eqnarray}
\mathcal{F}\left(t_{l_1}, \ldots, t_{l_L}\right) & =&\left(\sum_{m, m^{\prime}=\left\{k_1, \ldots, k_K\right\}} I_{m m^{\prime}}^* \prod_{n=\left\{l_1, \ldots, l_L\right\}} \phi_{m^{\prime} n}\left(t_n\right)\right)\left(\sum_{p, p^{\prime}=\left\{k_1, \ldots, k_K\right\}} I_{p p^{\prime}} \prod_{n^{\prime}=\left\{l_1, \ldots, l_L\right\}} \phi_{p^{\prime} n^{\prime}}^*\left(t_{n^{\prime}}\right)\right)\nonumber\\
& =&\sum_{m, m^{\prime}, p, p^{\prime}=\left\{k_1, \ldots, k_K\right\}} I_{m m^{\prime}}^* I_{p p^{\prime}} \prod_{n, n^{\prime}=\left\{l_1, \ldots, l_L\right\}} \phi_{m^{\prime} n}\left(t_n\right) \phi_{p^{\prime} n^{\prime}}^*\left(t_{n^{\prime}}\right)\nonumber\\
& =&\sum_{m, m^{\prime}, p, p^{\prime}=\left\{k_1, \ldots, k_K\right\}} I_{m m^{\prime}}^* I_{p p^{\prime}} \prod_{n=\left\{l_1, \ldots, l_L\right\}} \phi_{m^{\prime} n}\left(t_n\right) \phi_{p^{\prime} n}^*\left(t_n\right).
\end{eqnarray}
Integrating $\mathcal{F}(t_{l_1},...,t_{l_L})$ with respect to $t_{l_1},...,t_{l_L}$, we get a time-averaged fidelity, 
\begin{eqnarray}
        \overline{\mathcal{F}}&=&\int \cdots \int d t_{l_1} \cdots d t_{l_L} \mathcal{F}\left(t_{l_1}, \ldots, t_{l_L}\right)\nonumber\\
        &=&\sum_{m, m^{\prime}, p, p^{\prime}=\left\{k_1, \ldots, k_K\right\}} I_{m m^{\prime}}^* I_{p p^{\prime}} \prod_{n=\left\{l_1, \ldots, l_L\right\}} \int \phi_{m^{\prime} n}\left(t_n\right) \phi_{p^{\prime} n}^*\left(t_n\right) d t_n.
\end{eqnarray}
\end{widetext}

Since $\int \phi_{m^{\prime} n}\left(t_n\right) \phi_{p^{\prime} n}^*\left(t_n\right) d t_n$ is given by the first-order coherence function between the individual
atomic single-photon sources, the time-averaged fidelity can be obtained only form the first-order
coherence functions.

\bibliography{ref}

\end{document}